\documentclass[aps,prl,twocolumn,floats,superscriptaddress]{revtex4}
\usepackage{amsmath}
\usepackage{bm}
\usepackage{graphicx}
\usepackage{epstopdf}
\usepackage{color}
\newcommand{\nc}{\newcommand}
\nc{\rr}{\color{red}}
\nc{\bb}{\color{blue}}
\nc{\gr}{\color{green}}
 
\begin{document}
 
\title{Inflationary dynamics for matrix eigenvalue problems }
\author{Eric J. Heller}
\affiliation{Department of Physics, Harvard University, Cambridge, MA 02138, USA}
\affiliation{Department of Chemistry and Chemical Biology, Harvard University, Cambridge, MA 02138, USA }
\author{Lev Kaplan}
\affiliation{ Department of Physics, Tulane University, New Orleans,
LA 70118, USA}
\author{Frank Pollmann}
\affiliation{ Max-Planck-Institute for the Physics of Complex
Systems, D-01187 Dresden, Germany}

\date{December 25, 2007}

\begin{abstract}
Many fields of science  and engineering require finding   eigenvalues and eigenvectors of large matrices. The solutions can represent oscillatory modes of a bridge,  a  violin, the disposition of electrons around  an atom or molecule,   the acoustic modes of a concert hall, or hundreds of other physical quantities.  Often only the few eigenpairs with the lowest or highest frequency (extremal solutions) are needed.  Methods that have been developed over the past 60 years   to solve such problems include the Lanczos~\cite{lanczos,paige} algorithm, Jacobi-Davidson techniques~\cite{davidson}, and the conjugate gradient method~\cite{lls}. Here we present a way to solve the extremal eigenvalue/eigenvector problem, turning it into a  nonlinear classical mechanical system  with a modified Lagrangian constraint.  The constraint induces exponential  inflationary growth of the desired extremal solutions.
\end{abstract}

\maketitle
Physical problems of importance to many fields of science   are routinely reduced to an eigenvalue problem for a real symmetric, or hermitian, $N \times N$ matrix $A$:
\begin{equation}
A \psi_n = e_n  \psi_n
\end{equation}
where $\psi_n$ is the $n^{th}$ eigenvector with eigenvalue $ e_n$.  
Realistic simulations  can generate matrices of dimension $N=10^9$ or more, but often most of the matrix elements vanish for physical reasons,  yielding a sparse matrix.  
 Well established methods exist, which focus on finding extremal eigenpairs (or internal eigenpairs with various pre-conditioning strategies), such as conjugate gradient methods~\cite{lls}, Jacobi-Davidson techniques~\cite{davidson}, and Lanczos algorithms~\cite{lanczos,paige,RRGM}. 
 
 The  bulk of the numerical effort in these approaches goes into the iteration step, e.g., multiplying the matrix $A$ by a vector $\phi$. The methods differ as to how information from each iteration is used. Lanczos and Arnoldi methods use a Krylov space spanned by the initial (often random) vector  $\phi_1$ and its iterates $\phi_n = A \phi_{n-1}$.  
 
 Within the class of Krylov space approaches, diagonalizing in the full Krylov space  is  variationally optimal and in principle cannot be bested by  any other use of the Krylov vectors, such as the one we propose here.   However, if this were the end of the story there would be little need for restarting algorithms (such as implicitly re-started Arnoldi, used in the ARPACK library and in the MATLAB environment for example), or for the many other strategies that have been proposed.  One reason for these strategies, and the continued activity in the field, is that in practice memory issues (in storing the matrix $A$ and/or the Krylov vectors) and numerical stability issues both limit the performance of the ideal Lanczos method. 
 
  Furthermore, the subject of optimal approaches to large matrix eigenvalue problems remains active due to special requirements associated with different problems (such as the need for interior eigenpairs, the number of eigenpairs needed, the accuracy required, etc.), and the existence of classes of matrices with special convergence characteristics (diagonal dominance, large or small spread of diagonal elements, near degeneracy of lowest eigenvalues, etc.).
 Non-Krylov space approaches, especially the Jacobi-Davidson method, have been invented to address some of these issues.  The Davidson method, which uses pre-conditioning, has been invaluable for quantum chemistry applications, especially where several lowest eigenpairs are needed, the spread of diagonal elements is large, and the matrices, though sparse, still contain too many nonzero elements to store.
 
Our purpose in this paper is to report an approach that starts from a fresh premise. Although it also relies on matrix-vector multiplication and is not immune to the issues that limit the standard approaches, it is different enough in its design and  implementation to deserve special attention. The main idea is to  replace  a  large real symmetric (or hermitian) $N$-dimensional eigenvalue problem  with an $N$-dimensional classical trajectory problem, where the potential energy minimum corresponds to the minimum eigenvalue $e_0$ and the coordinates of this minimum correspond to the eigenvector $\psi_0$.
Even with a billion dimensions, for the proper choice of parameters one quickly finds the minimum under ``time'' evolution (discrete iterations).  
 There are no false local minima. We show that under this dynamics, the lowest eigenvectors grow exponentially fast relative to their neighbors, through a Lagrange multiplier that regulates this inflation. Nearby eigenpairs are also easily found.   
 Implementation of the algorithm is  simple, flexible, and robust, and convergence rates are easily proved analytically. 

Associating various kinds of eigenvalue problems with dynamical systems is not new; it is especially popular in the context of quantum mechanics.    Examples   include the Pechukas approach to random matrix eigenvalue spectra~\cite{pechukas},   the Miller-Meyer-McCurdy classical analog for electronic degrees of freedom~\cite{mmm}, the powerful and popular idea of replacing quantum statistical mechanics with a classical polymer bead system~\cite{cw}, and the well known association of classical driven, damped oscillators with quantum transitions in spectroscopy~\cite{driven}. We single out  the Car-Parrinello (CP) method and especially Car's damped CP method, proposed in the context of density functional theory (DFT), as most closely related to the present work~\cite{car}.  However we treat here the general real symmetric (or hermitian) eigenvalue problem; there is no connection to DFT, the Kohn-Sham equations, or even quantum systems.  Nonetheless, the potential is there in the future to try to construct on the fly CP-like methods that are not DFT based.  
   
The real symmetric eigenvalue problem is equivalent to finding the principal axes or ``normal modes'' of the harmonic potential  defined by 
$
A(\vec x) =  \sum_{i,j} x_i A_{i,j} x_j 
$,
where the $x_i$ are thought of as real orthogonal coordinates (we will not explicitly write down expressions for the hermitian case, but the extension is trivial).  We adopt a Lagrangian approach initially, but several modifications to follow may take us away from a strict Lagrangian dynamics. 
We take  the Lagrangian to be
\begin{equation} 
{\cal L} =   \sum\limits_i \dot x_i^2 -   \sum\limits_{i,j} x_i A_{i,j} x_j  + \lambda  \, \left(\sum_i x_i^2 -1 \right) \,,
\end{equation}
where the Lagrange multiplier $\lambda   $ enforces normalization.  Lagrange's equations require   $   \lambda(t) =  A(\vec x(t)) = \sum_{i,j} x_i(t) A_{i,j} x_j(t) $, i.e., the potential energy at time $t$. The Euler-Lagrange equations read
\begin{equation}
\label{govern2}
\ddot x_i = -   \sum\limits_j A_{i,j} x_j +   \lambda(t) x_i ; \ \ 1 \le i \le N \,.
\end{equation}

Defining $p_i =  \dot x_i$, and applying a naive Euler integrator with time step $\delta t$ gives the discrete-time mapping
 \begin{eqnarray} 
 \label{notnormal}
p_i(t+ \delta t) &=& p_i(t)- \sum\limits_j \left [A_{i,j}-\lambda(t)\delta_{ij}\right ] x_j(t) \,\delta t \cr
x_i(t+ \delta t) &=& x_i(t)+  p_i(t+\delta t) \,\delta t \,.
 \end{eqnarray} 
(More sophisticated discretizations, such as Verlet, also work well.) It is revealing  to make a linear transformation  to the (as yet unknown)  normal (eigenvector) momenta and coordinates $\pi_i, \xi_i$:
\begin{eqnarray} 
\label{normal}
\pi_i(t+ \delta t) &=& \pi_i(t) -  [e_i-\lambda(t)] \xi_i(t) \,\delta t \cr
\xi_i(t+ \delta t) &=& \xi_i(t)+  \pi_i(t+\delta t) \,\delta t \,.
 \end{eqnarray} 
Iterating Eq.~\ref{notnormal} is mathematically equivalent to iterating Eq.~\ref{normal}, which is a discrete area-preserving map corresponding to a set of $N$ independent damped harmonic oscillators.

For sufficiently small time step (not necessarily the regime we want to be in numerically) and assuming temporarily that $\lambda(t)$
is constant, we have 
\begin{equation}
\label{notinflating}
\xi_i(t)  =a_i  \cos( \omega_i t +\delta_i) 
 \end{equation}
   if $e_i > \lambda $, where $\omega_i  = \sqrt {e_i-\lambda} $   and $\delta_i$ is a real 
  phase shift (which along with $a_i$ depends on initial conditions
  $\xi_i(0)$ and $\pi_i(0)$). On the other hand, for low eigenvalues 
    $e_i < \lambda $, the normal coordinates evolve as  
 \begin{equation}
 \label{inflating}
\xi_i(t)  = (a_i' e^{ \omega_i^\prime t}+b_i' e^{-\omega_i^\prime t})\,,
 \end{equation}
  where $\omega_i^\prime  = \sqrt {\lambda -e_i} $,
  and the first term obviously dominates at long times.
Of course $\lambda(t)$ is not constant in practice, but it does quickly become slowly decreasing.  Thus we see 
that eigenmodes with eigenvalues $e_i$ below
  $\lambda(t)$ 
 (which set always includes the ground state, by the variational theorem) are exponentially inflating at any given time $t$, while the higher modes become simple oscillators. We will say more about the optimal choice of $\delta t$ 
  below.  

The normalization $\sum_i x_i^2=1$ can no longer be enforced by the Lagrange multiplier $\lambda(t)$ when finite, and possibly large, time steps  $\delta t$ are taken. That is not a problem, because the vector norm is easily imposed numerically before each time step, or even at the very end of the calculation (since we may easily generalize the previous expression for $\lambda$ to $\lambda=\sum_{i,j} x_i A_{i,j}x_j/\sum_i x_i^2$, making Eqs.~\ref{notnormal} and \ref{normal} equally valid for $\vec x$ of any norm). However $\lambda(t)$ retains the more important job of regulating  inflation, by controlling the border between inflating and non-inflating states. This is a key point. From this point of view, there is no reason for $\lambda(t)$ in Eq.~\ref{notnormal} or \ref{normal} to be strictly defined as the current potential energy estimate at time $t$.  Instead, we may replace it with $\tilde \lambda(t)$, a time-dependent parameter under our control to help optimize convergence by inflating the desired eigenmodes as rapidly as possible relative to the other modes. One may show, through a simple rescaling of variables, that this replacement is mathematically equivalent to adding a time-dependent damping term to Eq.~\ref{govern2}: $\ddot x_i=\cdots-\gamma(t) \dot x_i(t)$, where $\tilde \lambda=\lambda+\gamma^2/4$.

To calculate the rate of convergence of the inflation method, it is sufficient to consider inflation of the ground state coordinate $\xi_0$ relative to the first excited state coordinate $\xi_1$, since ground state inflation relative to higher excited states is obviously at least as fast. From Eqs.~\ref{notinflating} and \ref{inflating}, we have
 $\xi_0(t)/\xi_1(t) \sim e^{{\rm Re}(\sqrt{\lambda-e_0}-\sqrt{\lambda-e_1})t}$.  
 The exponent is peaked, and thus the ground state is approached fastest, when the inflation border
    $\lambda$ 
  is precisely equal to $e_1$, the eigenvalue of the first excited state. In that case the ground state coordinate grows relative to every other at the fastest possible rate,
\begin{equation}
\label{conticonvergence}
\xi_0(t)/\xi_n(t) \sim e^{\sqrt{e_1-e_0}\,t} \,.
\end{equation}

In practice  we do not know the energy $e_1$ a priori, but given an estimate of the gap $\epsilon_{01} = e_1-e_0 $, which can be obtained in several ways, we can set 
$\tilde \lambda(t) = \lambda(t) + \epsilon_{01} $, 
ensuring that as $\lambda(t)$ approaches the ground state energy $e_0$ we get maximal possible inflation of the ground state.  When working with a class of similar physical systems, the simplest approach is to use the typical value of the gap as our initial estimate for $\epsilon_{01}$, and set the initial value of $\tilde \lambda$ accordingly. One can do much better in a particular case by performing some number of iterations using an initial guess for $\epsilon_{01}$ (to cleanse the estimate of high lying eigenpairs), and then varying this guess, while noting which value gives the steepest descent of $\mu = {\rm Tr} \left [(A - \lambda(t))^2\right ]$, which is a standard measure of error that requires no
additional matrix-vector multiplications. The optimal estimate of the gap will result in the error decaying as $\mu \sim e^{-2\epsilon_{01}t}$, providing an obvious consistency check on our estimate.

The convergence of the algorithm is limited by the time step $\delta t$, which we would like to take as large as possible to reach large times quickly. Strict accuracy is not a concern, since we  are only interested in inflating the lowest few modes relative to the others, not in faithful integration of the second order differential equations. However, too large a time step will destabilize the stable oscillators corresponding to very high eigenvalues (very large $e_i -\lambda$ in Eq.~\ref{normal}), causing inflation of the wrong modes. A short analysis shows that Eq.~\ref{normal} remains stable for large $e_i$ only if the time step $\delta t$ obeys 
$ \delta t < 2 / \omega_{\rm max}$, 
where $\omega_{\rm max}^2=e_{\rm max}-e_0$ is an upper bound on the possible values of $e_i-\lambda$. 
  Thus we have an approximate bound $\delta t < 2 / 
 \omega_{\rm max}$, which we have found works rather well in practice. Combining this result with the optimal rate of convergence in continuous time (Eq.~\ref{conticonvergence}), we find that the number of discrete steps required for convergence scales as
\begin{equation}
\label{numsteps}
\frac{t}{\delta t} \sim \sqrt{\frac{e_{\rm max}-e_0}{e_1-e_0}} \,.
\end{equation}

Of course to choose an appropriate time step $\delta t$ in a particular calculation, we need a rough estimate of the spectral range $e_{\rm max}-e_0$. In this regard, it is interesting to note that starting off with too large a time step does not ruin the calculation, as is revealed by continuing with the iterations at a smaller time step: the error $\mu(t)$ often quickly recovers, dropping dramatically in just a few steps.  This is easily understood from our earlier analysis: Too large a step may inflate some of the highest lying eigenpairs, with eigenvalues $e_i \sim e_{\rm max}$, ruining the measure $\mu$. However, these eigenpairs are precisely the ones  killed most rapidly as soon as the time step is reduced to the stable region. That is, inflation has already been working well over most of the spectrum, but this was not reflected in the error measure; the errant eigenpairs are easily eliminated once the time step is reduced.

The square root in Eq.~\ref{numsteps} is a result of using a second order differential equation in Eq.~\ref{govern2}. 
It is instructive to consider the more straightforward idea  of applying $\exp(-\beta A)$ to a trial vector $\phi(0)$, i.e., solving the first order equation $ d\phi(\beta) /d\beta = - A \phi (\beta)$. In a specific basis, this gives 
$ d x_i(\beta)/d\beta = -\sum_j A_{i,j} x_j(\beta)$, which can be discretized in (imaginary) time $\beta$. This very simple idea works of course, but only for much smaller time steps and with slower convergence. To prevent runaway growth of the high-lying eigenmodes, the time step $\delta \beta$ must be chosen so that $\delta \beta \sim 1/(e_{\rm max}-e_0)$, as compared with $\delta t \sim 1/\sqrt{e_{\rm max}-e_0}$ in the Lagrangian case, and the number of steps needed for convergence scales as $\beta /\delta \beta \sim (e_{\rm max}-e_0)/(e_1-e_0)$, i.e., the square of the number of time steps needed in the Lagrangian method (Eq.~\ref{numsteps}). The same speedup was found for the second order damped Car-Parrinello method~\cite{car}, and also applies to the conjugate gradient method.

It is tempting to try differential equations of even higher order $m \ge 3$,
e.g., $d^m x_i/dt^m =-\sum\limits_j A_{i,j} x_j + \cdots$, but this does not work, as it is impossible for all $m$ roots $\omega_i \sim (e_i+\cdots)^{1/m}$ to be in the same half-plane, for either sign of $(e_i+\cdots)$, and thus inflation will always occur both for large and small $e_i$.

\begin{figure*}[thb]
 \begin{center}
      \begin{tabular}{cc}
	(a)\includegraphics[height=6.3cm]{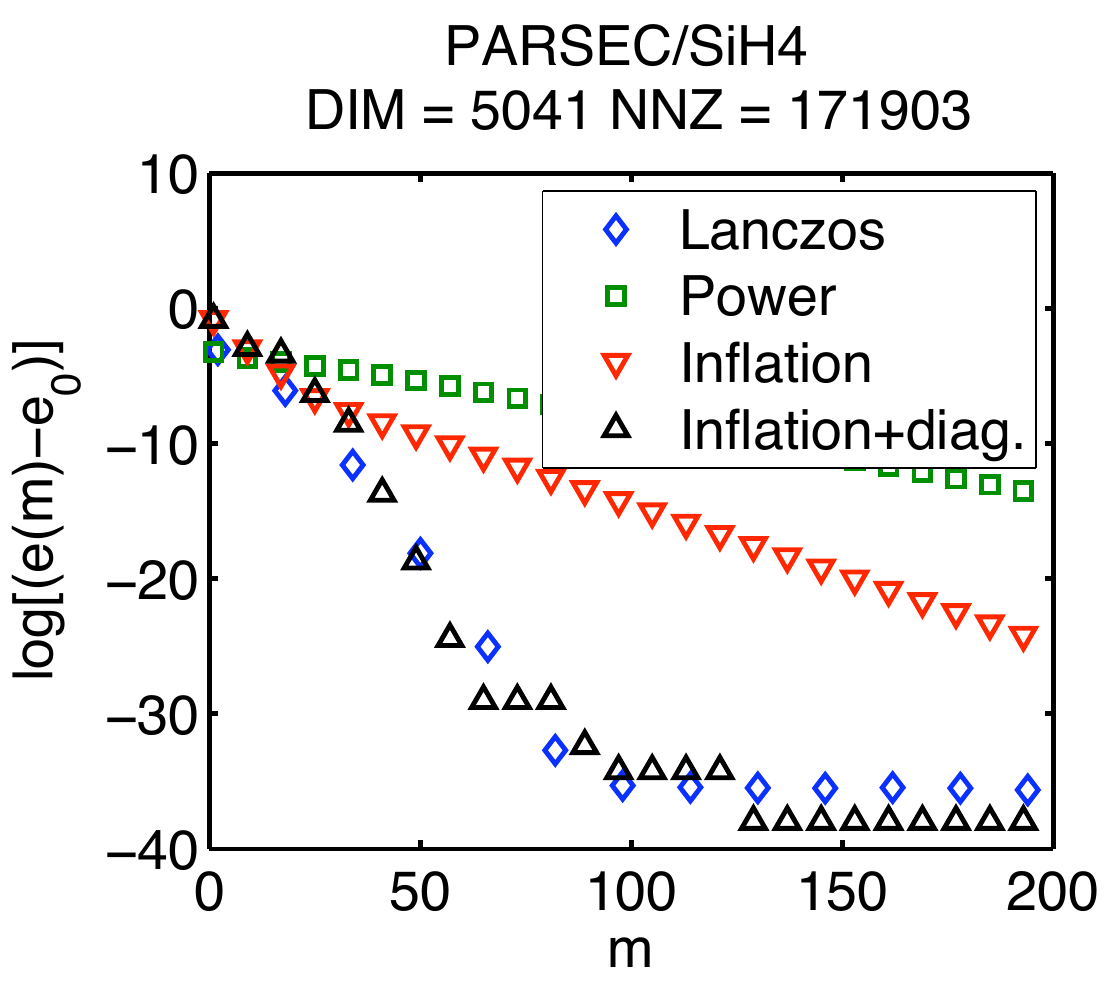}&
	(b)\includegraphics[height=6.3cm]{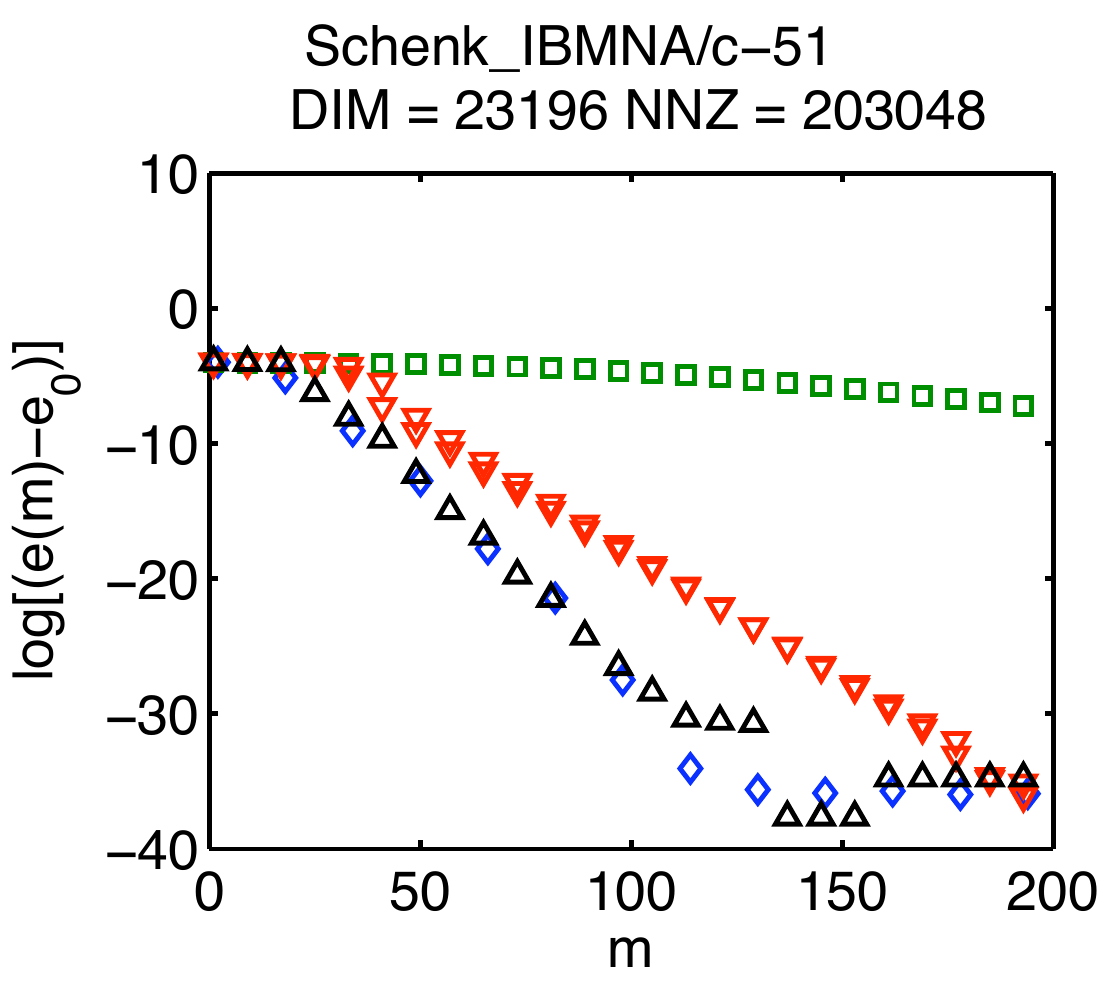}\\
	(c)\includegraphics[height=6.3cm]{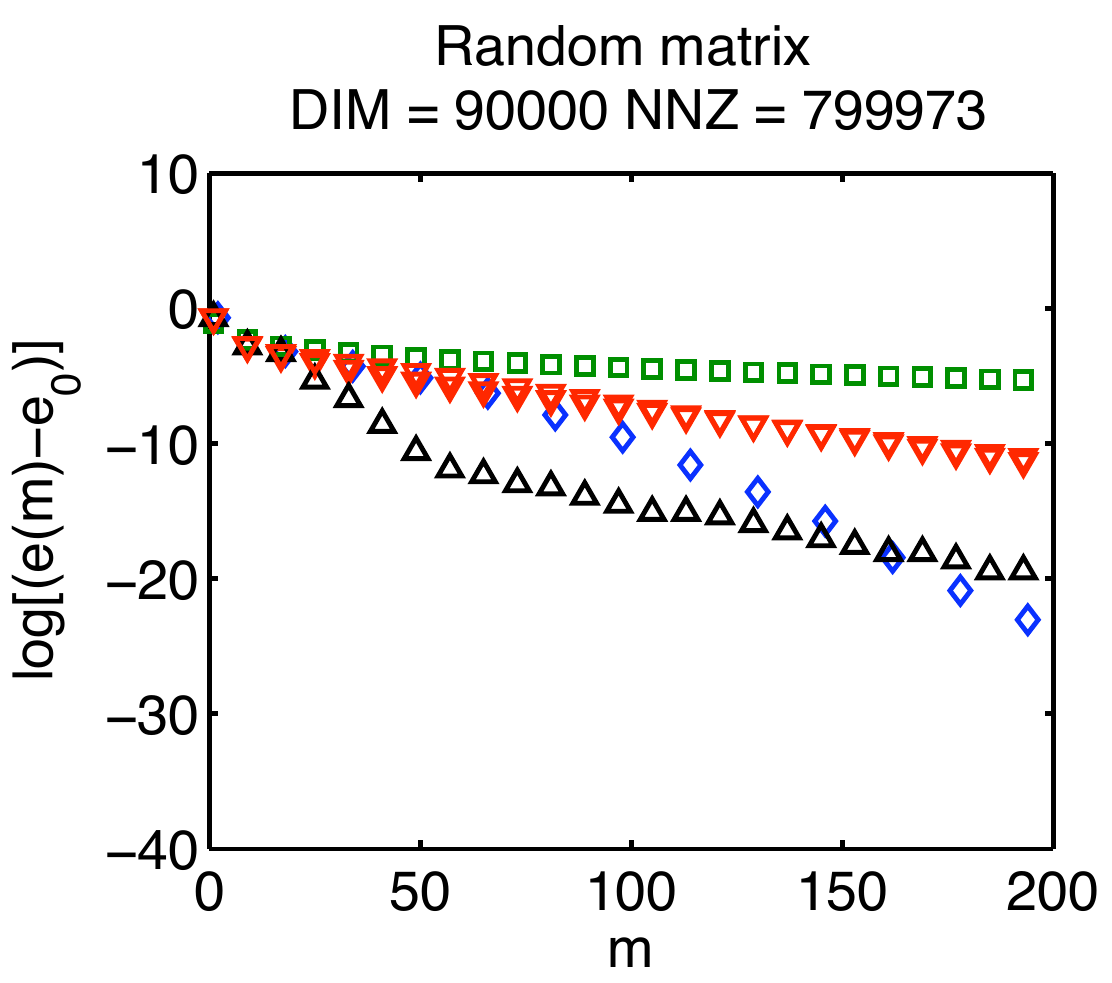}&
	(d)\includegraphics[height=6.3cm]{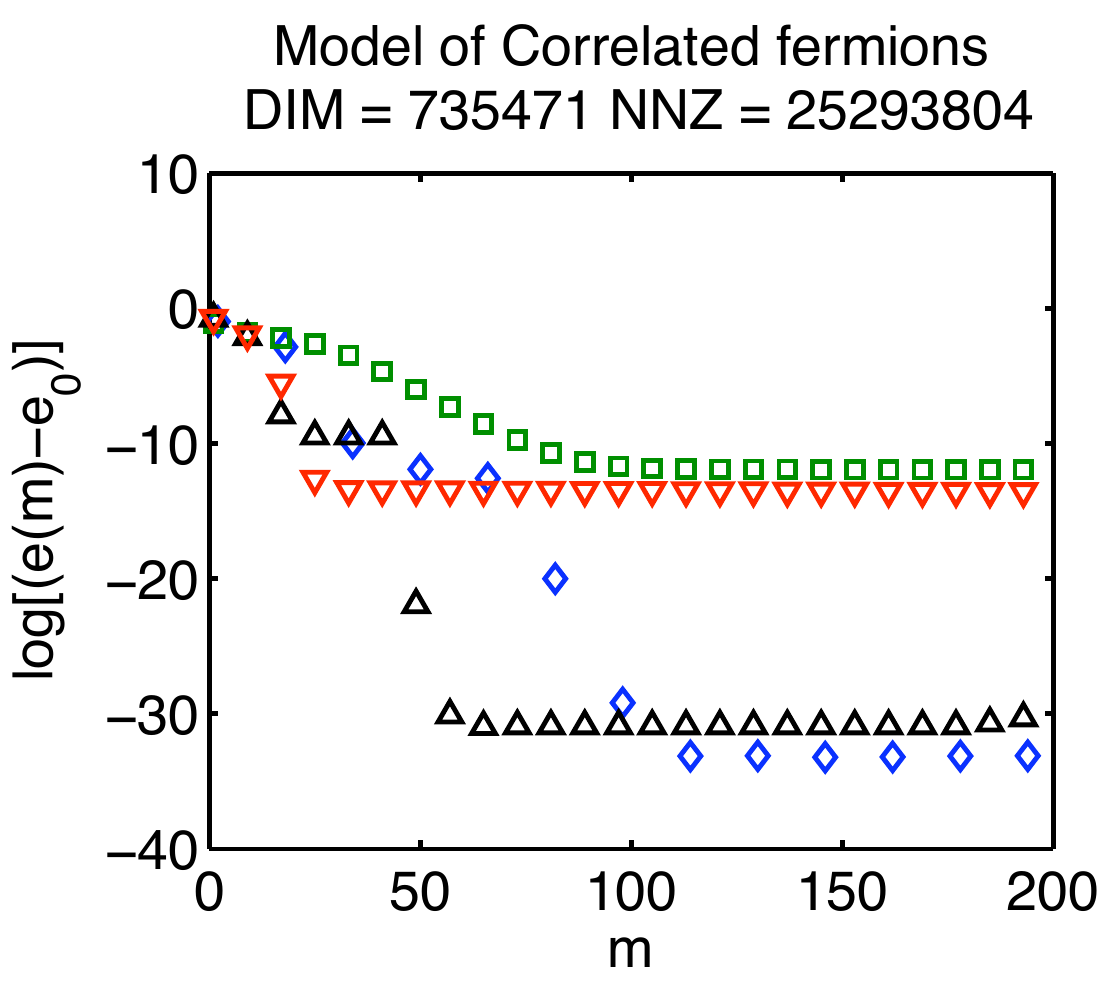}
      \end{tabular}
    \end{center}
    \caption{Comparison of convergence of the present inflation method with the Lanczos and Power methods. The computational time $m$ (to calculate both the lowest eigenvalue and corresponding eigenvector) is the number of matrix-vector multiplications. Note that the original Lanczos method requires two matrix-vector multiplications per step if the eigenstate as well as the eigenvalue are to obtained at the end of the calculation without storing all intermediate vectors. The implicitly restarted Arnoldi method (MATLAB/ARPACK) behaves very similarly to Lanczos, and is not shown. Panels (a) and (b) show results for some test matrices taken from \cite{uf_sparse_matrix} and (c) for a random sparse matrix. The matrix used in (d) corresponds to a model of strongly correlated spin polarized  fermions on a triangular lattice.  }	
\label{comparison}
\end{figure*}

The above analysis uncovers a  problem, common to iterative eigenpair methods: slow convergence to the ground state if one or more excited state energies are very close to the ground state energy. A very satisfactory solution exists for this problem in the present context.  Instead of waiting for inflation to separate out the ground state from these low-lying excited states, we admit the low-lying excited states into our calculation by choosing a window size $w$ and performing inflation with 
$\tilde \lambda(t)=\lambda(t)+w$. 
This choice optimally inflates away all modes with energy $e_i>e_0+w$, requiring only $\sim \sqrt{(e_{\rm max}-e_0)/w}$ steps for convergence, and the resulting vector $\phi$ contains (to any desired accuracy) only contributions from the $k$ states within the energy window $[e_0,e_0+w]$. Subsequently, $\phi$ is iterated an additional $k-1$ times, saving vectors $\phi_n$ and $A\phi_n$ after each iteration, and finally the hamiltonian $A$ is constructed and diagonalized explicitly in the subspace spanned by $\phi_0 \ldots \phi_{k-1}$. The parameter $k$ may be incremented until convergence to the lowest eigenpair is achieved. We note that the diagonalization is numerically trivial for moderate $k$, so the only significant additional cost is that associated with the storage of the $2k$ vectors $\phi_n$ and $A\phi_n$. A tradeoff between time and storage constraints determines the optimal window size $w$, as a larger $w$ requires fewer iteration steps, but makes it necessary for a greater number of vectors to be simultaneously stored in memory before the final diagonalization. We have successfully used this approach on various matrices (see below for details).

Several obvious generalizations can be implemented. The governing equations of the inflation method may easily be extended to non-orthogonal basis vectors. Also, the eigenvectors associated with the first excited state, second excited state, and so on, can easily be found by evolving several vectors simultaneously and including constraints that enforce their orthogonality to one another, i.e., $\vec \phi^\alpha\cdot \vec \phi^\beta = 0$ for $\alpha \ne \beta$. For example, this may be accomplished by adding a term $\nu_{\alpha\beta} \sum_i x_i^\alpha x_i^\beta$ to the Lagrangian, where $\nu_{\alpha\beta}$ is a Lagrange multiplier that can be shown  to equal $\sum_{ij} x_i^\alpha A_{i,j} x_j^\beta$. Alternatively, one may begin by dynamically evolving a single random vector using an appropriate window $w$, where $w$ is chosen to include all eigenvalues of interest. After eigenpairs lying outside the window have been inflated away, one performs $k \times k$ diagonalization using $k$ iterates of the initial vector as discussed above to obtain approximations for $k$ lowest-lying eigenpairs. The $k$ approximate vectors may then be dynamically evolved individually while enforcing the orthogonality constraints to obtain any desired accuracy for each eigenpair.

We have applied the inflation method to a considerable variety of large matrices, including diagonally dominated sparse, random sparse, and full, with diagonal elements chosen unevenly or evenly spaced, and eigenvectors weakly, moderately, or strongly mixed, up to  a size of $10^9 \times 10^9$ (see Fig.~\ref{comparison} for a selection). The limiting factor is not the dimension of the matrix {\it per se}, but rather the number $N_{nz}$ of its nonzero elements, both in terms of storage and time required for an iteration, which both scale linearly with $N_{nz}$.  These traits are also common to all methods employing matrix-vector multiplication.  The advantage of the above described diagonalization in the subspace spanned by $\phi_0 \ldots \phi_{k-1}$ is most visible in the model of spin polarized electrons on a triangular lattice (see Fig.~\ref{comparison}(d)).  This is a numerically very difficult problem because it has a high density of low-lying excitations. The low-lying states can be efficiently separated from the ground state by the diagonalization, leading to considerably better convergence.

\begin{figure}
\centerline {
\includegraphics[height=6.3cm]{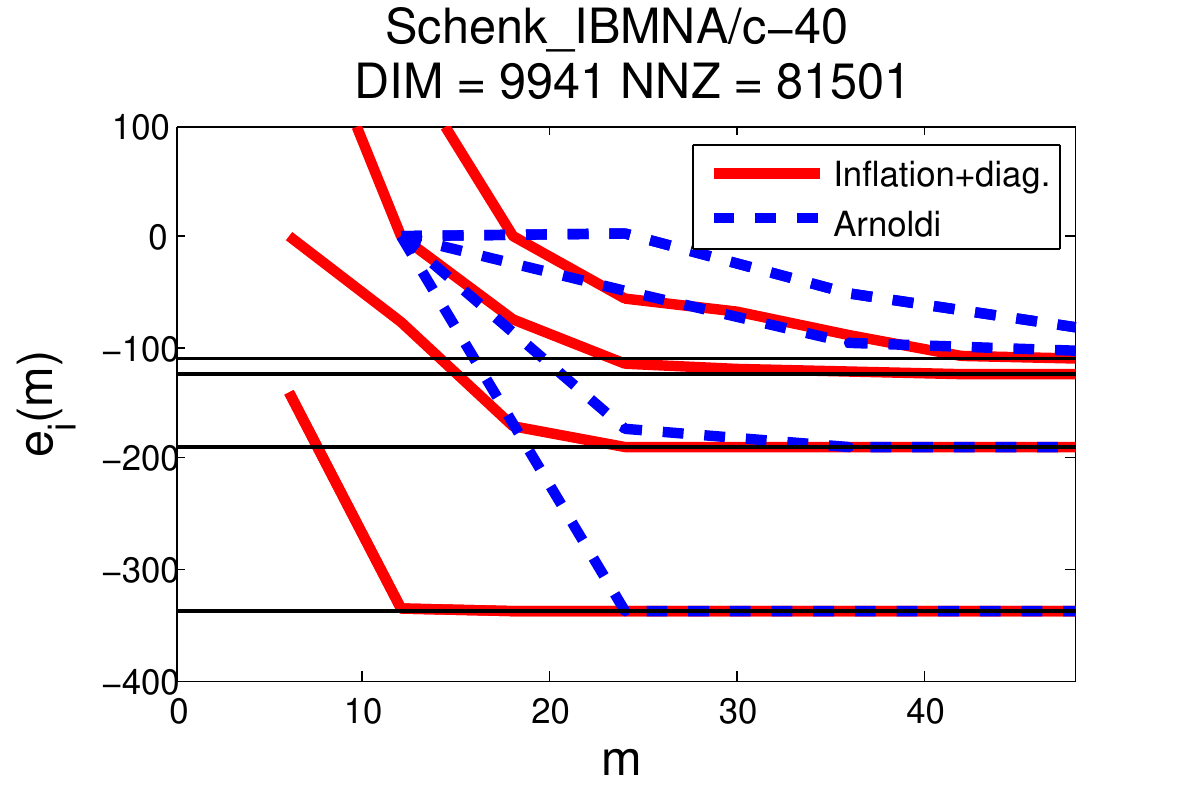}
}
\caption{The convergence of the inflation method for the lowest four eigenpairs of a test matrix~\cite{uf_sparse_matrix} is compared with
the implicitly restarted Arnoldi method (as implemented in MATLAB/ARPACK). Exact eigenvalues are indicated by horizontal lines. In the inflation method, we diagonalize in a 6-dimensional basis after every 6 dynamical steps. In the Arnoldi calculation, we use a basis of size $12$. In each case, the computational time $m$ represents the number of matrix-vector multiplications.}
\label{figmulti}
\end{figure}

In Fig.~\ref{figmulti} we show that multiple eigenpairs can be obtained simultaneously by dynamically evolving a single initial vector. Here we perform $6 \times 6$ diagonalization after every $6$ steps, and show convergence of the first $4$ eigenpairs. An arbitrary number of extremal eigenpairs can be obtained by diagonalizing in a subspace generated by inflating several eigenpairs under an orthogonality constraint.  

\begin{figure}
\centerline {
\includegraphics[width=3.5in]{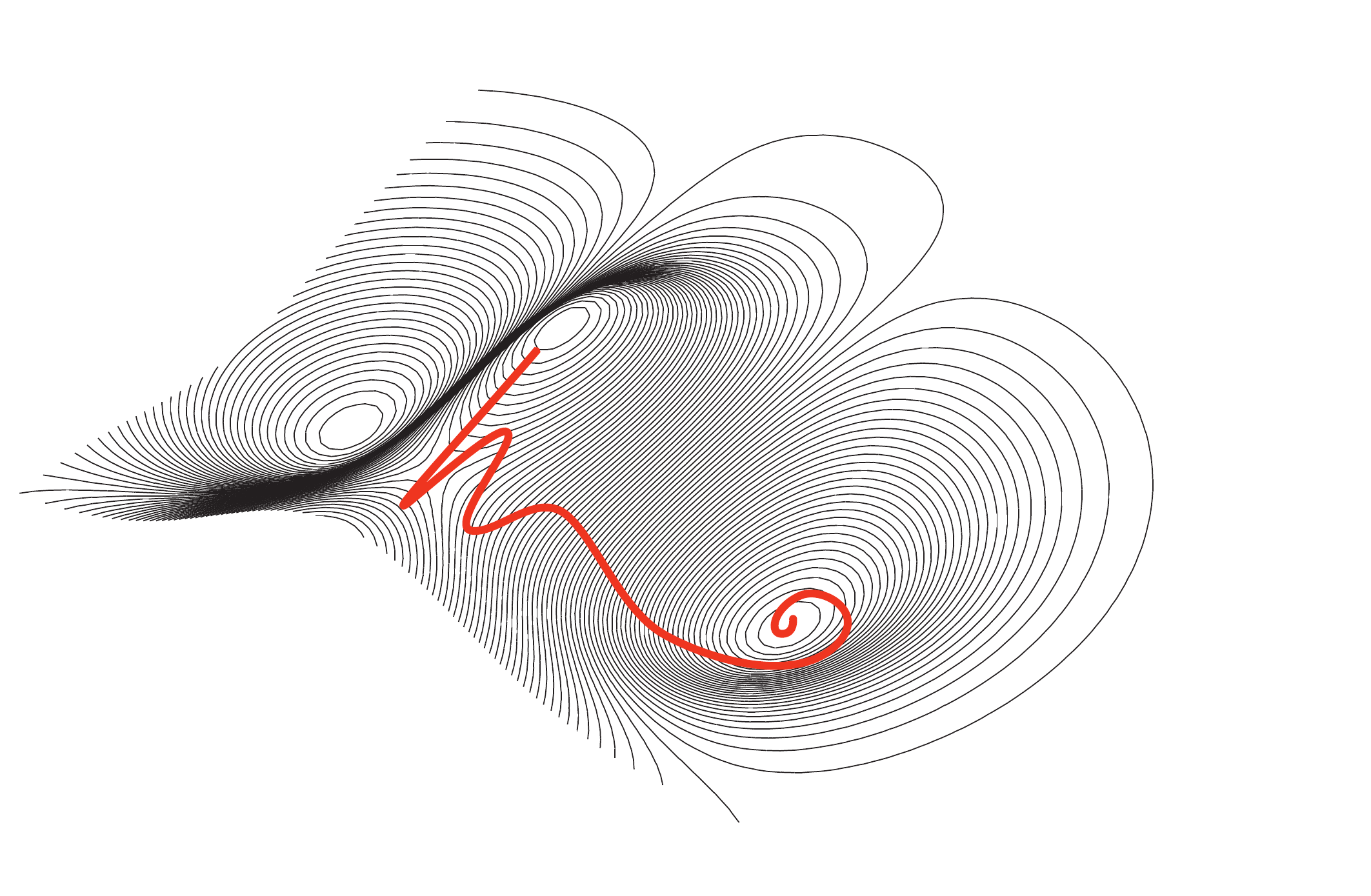}
}
\caption{An arbitrary starting vector  finds the global minimum in the pseudopotential, falling off any saddles it encounters, associated with eigenvalues other than the lowest one. In this diagram the two minima are topologically equivalent, corresponding to the same eigenvector with opposite sign. Thus it does not matter which well the trajectory finds.}
\label{topo}
\end{figure}

We have investigated a few significant variations of the ideas presented here and have found thus far that the inflation approach works best.  For example, the idea of using a dynamical system to find eigenpairs suggests the following alternative idea: Instead of a Lagrangian constraint, consider a ``soft'' normalization constraint, imposed by adding a smooth quartic potential term,
making the complete pseudopotential 
\begin{equation}
V  = \sum\limits_{i,j} x_i A_{i,j} x_j +  {\kappa}  \left (\sum\limits_i x_i^2 -1\right )^2.
\end{equation}
Again damped dynamics is used. In the limit of large $\kappa$, the quartic potential enforces unit norm of the solution vector, just as the Lagrangian constraint does.
Now consider the nature of the extrema of $V$. With no loss of generality, we make an orthogonal transformation to the (unknown) normal coordinates $\eta_i$,
\begin{equation}
V=   \sum\limits_{i} e_i \eta_i^2+  {\kappa} \left (\sum\limits_i \eta_i^2 -1\right )^2.
\end{equation}
The extrema of this potential are given (in the $\eta$ basis) by 
\begin{equation}
\frac {\partial V}{\partial \eta_i} = 2 \epsilon_i \eta_i + 4 \kappa  \left (\sum_j \eta_j^2 -1\right ) \eta_i = 0; \ \ {\rm all} \ i \,.
\end{equation}
Besides the trivial extremum at the origin (all $\eta_i=0$), we have up to $2N$ extrema corresponding to the $N$ possible eigenstates, where a single 
$
\eta_i = \pm \sqrt{1- e_i/2\kappa} \,
$
while $\eta_j=0$ for all $j \ne i$. For sufficiently strong constraint coefficient ($\kappa>e_0/2$), we easily check that all extrema are saddles with at least one unstable direction, with the exception of the extremum associated with the ground state (i.e., the global minimum). Thus, for almost all (i.e., all but a set of measure zero) initial conditions, 
the trajectory leads  downhill to the true minimum.  It does not dally long on any intermediate saddles it encounters   because of their exponential instability (see Fig.~\ref{topo}).  The global minimum is doubled to two equivalent solutions, which are related by prefactor of $-1$ and correspond to the same eigenstate.  It also does not matter if we adhere strictly to the rules of classical mechanics in getting to the minimum; any stepping method leading to the minimum will do.  
Note that the solution is independent of $\kappa$ and is exact, provided only that $\kappa>e_0/2$. 

The inflationary approach proposed here is quite competitive with   standard  methods.   We have not yet investigated pre-conditioning.   Although the inflationary method is quite general, it seems especially well suited to physics and quantum chemistry problems, because of its dynamical underpinnings. 

\section{Acknowledgements}  We thank Ernest Davidson for very helpful comments. This work was supported in part by the National Science Foundation under Grant No. PHY-0545390.
\bibliographystyle{unsrt}
\bibliography{my.bib}

 \end{document}